\documentclass[12pt]{article}

\usepackage{geometry}
\usepackage{amsmath}
\usepackage{amssymb}
\usepackage{graphicx}
\usepackage{pdflscape}
\usepackage{setspace}

\usepackage{color}
\usepackage{adjustbox}

\usepackage{amsmath}
\usepackage{graphicx}
\usepackage{subcaption}
\usepackage{pifont}
\usepackage{caption}
\usepackage{algorithm,algorithmic}
\usepackage{setspace}
\usepackage{color}
\usepackage{booktabs}
\usepackage[table,x11names]{xcolor}
\usepackage{multirow}

\usepackage[utf8]{inputenc}
\usepackage[english]{babel}

\usepackage[round,authoryear]{natbib}
\usepackage[T1]{fontenc}

\usepackage{verbatim}
\usepackage{soul}
\usepackage[title]{appendix}
\usepackage{booktabs,threeparttable}
\usepackage{cleveref}

\newcommand{\bfa} {\mbox{\boldmath $\alpha$}}

\newcommand{\bfO} {\mbox{\boldmath $\Omega$}}

\newcommand{\bfl} {\mbox{\boldmath $\lambda$}}

\newcommand{\bfmu} {\mbox{\boldmath $\mu$}}

\begin{document}

\newdimen\jot \jot=5mm
\newcommand{\fnote}[1]{\mbox{$\mbox{}^{#1}$}}
\newcommand{\bigsp}{\mbox{$\; \; \; \; \;$}}
\newcommand{\half}{\mbox{$\frac{1}{2}$}}
\newcommand{\bm}[1]{ \mbox{\boldmath $ #1 $} }
\begin{spacing}{1.2}
\begin{center}
{\bf \Large Integrative Bayesian Analysis of Brain Functional Networks Incorporating Anatomical Knowledge}\\~\\
Ixavier A. Higgins\\ \small Department of Biostatistics and Bioinformatics, Emory University, USA \\ Suprateek Kundu\\ Department of Biostatistics and Bioinformatics, Emory University, USA \\ Ying Guo\footnote{Research reported in this publication was supported by the National Institute Of Mental Health of the National Institutes of Health under Award Number ROI MH105561 and R01MH079448. The content is solely the responsibility of the authors and does not necessarily represent the official views of the National Institutes of Health.} \\ Department of Biostatistics and Bioinformatics, Emory University, USA
\end{center}
\end{spacing}
\vskip 12pt

\singlespacing
\begin{center}
{\noindent \bf Abstract } 
\end{center}
      Recently, there has been increased interest in fusing multimodal imaging to better understand brain organization by integrating information on both brain structure and function.  In particular, incorporating anatomical knowledge leads to desirable outcomes such as increased accuracy in brain network estimates and greater reproducibility of topological features across scanning sessions. Despite the clear advantages, major challenges persist in integrative analyses including an incomplete understanding of the structure-function relationship and inaccuracies in mapping anatomical structures due to inherent deficiencies in existing imaging technology.  This calls for the development of advanced network modeling tools that appropriately incorporate anatomical structure in constructing brain functional networks. We propose a hierarchical Bayesian Gaussian graphical modeling approach which models the brain functional networks via sparse precision matrices whose degree of edge-specific shrinkage is a random variable that is modeled using both anatomical structure and an independent baseline component. The proposed approach adaptively shrinks functional connections and flexibly identifies functional connections supported by structural connectivity knowledge.  This enables robust brain network estimation even in the presence of mis-specified anatomical knowledge, while accommodating heterogeneity in the structure-function relationship. We implement the approach via an efficient optimization algorithm which yields maximum a posteriori estimates. Extensive numerical studies involving multiple functional network structures reveal the clear advantages of the proposed approach over competing methods in accurately estimating brain functional connectivity, even when the anatomical knowledge is mis-specified up to a certain degree. An application of the approach to data from the Philadelphia Neurodevelopmental Cohort (PNC) study reveals gender based connectivity differences across multiple age groups, and higher reproducibility in the estimation of network metrics compared to alternative methods. \\


\newpage
\section{Introduction}
The human brain is an extremely complex organ responsible for all thought and bodily function.  Various approaches have sought to explain the brain's functionality as a result of neurotransmissions between individual neurons, reflected as the co-activation between voxels in brain images.  Recently, there has been a rapid increase in research on brain connectome analysis focused on linking inter-regional dependencies to brain function. Methods for both structural connectivity (SC) and functional connectivity (FC) have seen increasing developments with the emergence of non-invasive technologies such as diffusion tensor imaging (DTI) and functional magnetic resonance imaging (fMRI). FC measures the temporal coherence in brain activity across two distinct brain regions, while SC approaches based on DTI reconstruct white matter pathways in the brain by measuring the diffusivity of water molecules in brain tissues. These two types of connectivity offer complementary and interdependent information about brain structure and function.

Despite strong evidence regarding the role of white matter fiber tracts in regulating FC \citep{Damoiseaux2009, Honey2010, Sporns2013} and considerable progress in separately estimating FC and SC, there have been comparatively limited advances in FC approaches which are guided by underlying anatomical knowledge. Incorporating anatomical knowledge in estimating FC is clearly desirable since it is expected to produce more accurate estimates of the network, which translates to greater reproducibility of the findings as illustrated via our fMRI data analysis. However, several considerations need to be taken into account, such as the complexity of the structure-function relationship \citep{Hermundstad2013}, heterogeneity in FC for a given SC strength, which presumably is attributed to the fact that FC is only partially dependent on SC \citep{Damoiseaux2009, Messe2014} and regulated by unobservable dynamics in underlying neuronal activity \citep{Bressler2006}.

\begin{figure}[h!]
 \centering
\mbox{\includegraphics[height=3in,width=1\textwidth,keepaspectratio]{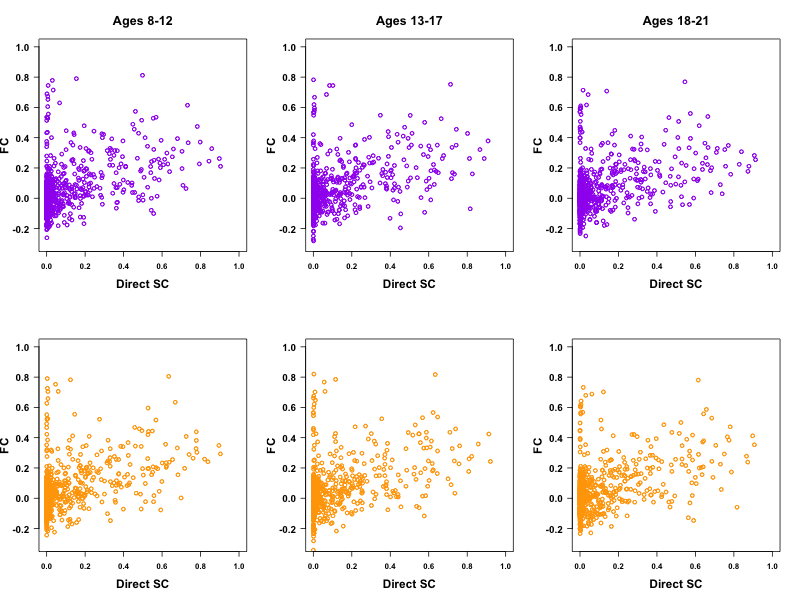}}
 \caption{\small Plots of the associations between partial correlation (FC) and direct structural connectivity (SC) for all males ({\it top row}) and females ({\it bottom row}) in our study.  The subjects fMRI data is pooled within each age and gender stratum while the representative anatomical connectivity is an average of all constituent SC information.  While mild positive correlations are observed between SC and FC, FC exists between regions with little to no direct structural connections.  We also observe high variability in FC for a given SC level.}
 \label{figure:FemalesSCFC}
\end{figure}

Recently, \cite{Venkataraman2012} and \cite{Xue2015} proposed approaches to jointly model the probability of co-activation based on fMRI data while incorporating direct structural connections.  They provide measures of functional co-activation deviating from standard measures of FC such as Pearson or partial correlations.  \cite{Hinne2014} proposed a Bayesian approach which uses fMRI data to model the distribution of partial correlations for edges determined by the given SC information. Hence, functional connections only exist between anatomically connected regions which is an extremely restrictive assumption that does not reflect a realistic relationship between brain structure and function.  Moreover, the above approaches use multi-subject data which requires registration of images to a shared template under the assumption that the volumes are similar and can be matched. Unfortunately, this assumption has limitations for human brain images considering the substantial variability in cortical anatomy and function \citep{Zhu2012}. This variability is especially pronounced during the developmental phases of childhood and adolescence as is the case with our motivating Philadelphia Neurodevelopmental Cohort (PNC) study. \cite{Ng2012} and \cite{Pineda2014} proposed approaches for estimating sparse functional networks for individual subjects via an adaptive graphical lasso involving edge specific shrinkage parameters which are parametric functions of the SC strengths. Under these approaches, FC with less anatomical support are more heavily penalized, and vice-versa. However, the parametric form of the shrinkage parameters may not adequately capture the complex underlying structure-function relationships and does not account for heterogeneity in FC for a given SC strength resulting from non-anatomical sources of variation (see Figure \ref{figure:FemalesSCFC} for a representative FC-SC relationship in the PNC study).  Moreover, such a parametric relationship may lead to erroneous FC estimates when the SC information is mis-specified, rendering these approaches sensitive to the quality of anatomical information.

The above discussions highlight a serious need for integrative modeling approaches which adaptively estimate FC by incorporating structural knowledge in an appropriate manner. In designing such an approach, our primary goals for the method are that it will (a) provide accurate and reproducible brain network estimation, which is a topic of great importance in current literature \citep{Varoquaux2010}; and (b) specify a flexible structure-function relationship which is robust to mis-specification of SC information (arising from limitations in existing image acquisition technology) and can accommodate heterogeneity in FC for a given SC. We propose a novel hierarchical Bayesian Gaussian graphical modeling approach for estimating FC based on single subject fMRI data which incorporates given SC information in a manner that addresses the aforementioned objectives. The FC is computed via sparse precision matrices whose elements are estimated under Laplace type priors having edge specific shrinkage parameters that are random variables modeled using SC information and an independent baseline component. The prior encourages stronger FC given a large SC (and vice-versa), but also accounts for edge specific variations in FC unrelated to the brain anatomy, via the baseline component. The approach is thus flexible in accounting for anatomical knowledge with the FC being guided by, but not completely determined by, the SC information. Our method is motivated by the variable selection approach presented in \cite{Chang2016} which incorporates prior graph knowledge in a linear regression setting, but is distinct in addressing graphical model selection and precision matrix estimation, as well as the manner in which the prior knowledge is incorporated. Under certain choices of model parameters, the proposed approach reduces to an adaptive shrinkage approach specifying a parametric relationship between the shrinkage parameters and the anatomical information, similar to \cite{Ng2012} and \cite{Pineda2014}.

While Markov chain Monte Carlo (MCMC) can be used to implement the proposed approach, it is not scalable to large networks needed for whole brain connectome analysis as in our application. Moreover, MCMC samples cannot take exact zeroes under a Laplace prior, and an additional thresholding step is often needed for model selection. We propose an optimization algorithm to obtain the {\it maximum a posteriori} (MAP) estimate, which is computationally efficient, scales to a large dimensions, and does not require post-hoc thresholding. Under various simulation studies, we observe superior performance of the proposed method as compared to alternative approaches with or without SC information.  The advantages of our approach become more evident as the degree of mis-specification of anatomical knowledge increases, and/or the number of nodes grows larger which is particularly relevant for whole brain connectome analysis. 

Our efforts are motivated by data from the PNC study, a large-scale, NIMH funded initiative to understand the developmental trajectory of the brain from childhood to adolescence \citep{Satterthwaite1}. The PNC data contains both DTI and resting state fMRI measurements from boys and girls ages 8-21 years, with suggestive but unclear SC-FC relationships (Figure \ref{figure:FemalesSCFC}). We fuse multimodal brain imaging data to examine gender differences in brain networks across three age brackets - pre-teens (ages 8-12), teens (ages 13-17), and young adults (ages 18-21) - and discover several gender based differences in FC within and across the age brackets. We also assess the reliability of computed network metrics across scanning sessions and find that the proposed approach yields strong reproducibility in the estimation of network metrics which is almost always higher than alternative approaches. While other studies have separately examined the reproducibility of functional and structural brain networks (see \cite{Welton2015} for a review), ours is one of the first to examine the reproducibility of anatomically informed functional networks to  the best of our knowledge.

Section 2 describes the proposed methodology and the optimization routine for estimating networks, while sections 3 presents numerical studies and application of our method to PNC data. We conclude with a brief discussion in Section 4.
	
\section{Materials and methods}
\subsection{Gaussian graphical model for brain networks}

While early work on brain network estimation utilized Pearson correlation to measure undirected inter-regional dependencies, recent literature has focused on Gaussian Graphical Models (GGMs) which estimate a sparse inverse covariance matrix and measure functional connections via partial correlations \citep{Smith2011}. GGMs assume observations are normally distributed and that zeros in the inverse covariance matrix correspond to absent edges in the network, $\mathcal{G}$. A typical GGM specifies $\small {\bf y}_{t} \sim N_p(0,\boldsymbol{\Omega}^{-1}), t=1,..,T,$ where ${\bf y}_t$ represents the $p$-dimensional pre-processed fMRI signal at time t across all $p$ ROIs, $\boldsymbol{\Omega}(\boldsymbol{\Sigma}^{-1})$ is the inverse covariance matrix (or the precision matrix) which belongs to the cone of $p\times p$ symmetric positive definite matrices, denoted by $M_p^{+}$. Under this framework, estimating $\mathcal{G}$ is equivalent to estimating structural zeros in the positive definite precision matrix $\Omega$.

Due to tradeoffs between the cost and efficiency of information transfer, it is typically assumed that the brain seeks efficient organization favoring a sparse set of active connections at any point in time \citep{Eavani2015}. The GGM approach is well equipped to handle such sparse networks by imposing penalties that shrink sufficiently weak functional connections to zero, where the $L_1$ penalty under the graphical lasso \citep{Friedman2007} is a popular choice \citep{Ng2012, Monti2014, Pineda2014}. The graphical lasso can be thought of as an extension of the Lasso approach in regression settings and penalizes the full data likelihood to estimate the inverse covariance matrix as
\begin{equation}\label{eqn:baseGGM}
\small \boldsymbol{\hat \Omega} = \arg \max_{\boldsymbol \Omega \in M^{+}_p} \quad log\lvert \boldsymbol \Omega \rvert -tr(S\boldsymbol \Omega) - \lambda \lvert \boldsymbol \Omega \rvert_1,
\end{equation}\\
 where $S=\sum_{t=1}^T ({\bf y}'_t {\bf y}_t)/T$ is the sample covariance matrix, $\lvert \boldsymbol \Omega \rvert_1$ denotes the element-wise $L_1$ norm, and $\lambda>0$ is the penalty parameter controlling overall network sparsity. If $\lambda=0$, one obtains the maximum likelihood estimate, while large values of $\lambda$ shrinks an increasing number of off-diagonal elements to zero. The typical graphical lasso approach fits a series of graphs under various choices of the tuning parameter $\lambda$ and chooses the optimal network as the one minimizing some goodness of fit criteria \citep{Yuan2006}.

\subsection{Structurally informed Bayesian Gaussian graphical model}
Bayesian GGM approaches have been successfully used for estimating brain networks (see \cite{Mumford2014} for a review). One such approach is the Bayesian graphical lasso, which has similarities with the penalized graphical lasso approach in the sense that the {\it maximum a posteriori} (MAP) estimator is equivalent to the estimate under (\ref{eqn:baseGGM})  (see \cite{Wang2012}). The Bayesian graphical lasso, with a common shrinkage parameter, is defined as follows
\begin{eqnarray}
\small {\bf y}_t \sim N_p(0,\boldsymbol{\Omega}^{-1}), \mbox{ } t=1,\ldots,T, \mbox{ } \pi(\boldsymbol{\Omega} \mid \lambda) \propto \prod_{k=1}^p Exp(\omega_{kk}; \lambda) \prod_{j<k} DE( \omega_{jk}; \lambda) I(\boldsymbol{\Omega}\in M_p^{+}), \label{eq:BayesGGM}
\end{eqnarray}
as in Wang et. al (2012), where the diagonal element $\omega_{kk}$ is modeled under an exponential prior distribution $Exp(\lambda)$, the off-diagonal element $\omega_{jk}$ is modeled with double exponential or Laplace prior distribution $DE(\lambda)$, $\lambda$ is the shrinkage parameter, and $I(\cdot)$ denotes the indicator function. In a fully Bayesian paradigm, $\lambda$ is typically assigned a prior distribution, and is thus learned from the data, resulting in an adaptive shrinkage of the elements in $\Omega$.

In order to incorporate anatomical knowledge in functional connectivity estimation, we propose a hierarchical Bayesian structurally informed Gaussian graphical model (siGGM).  It is based on the generic Bayesian GGM in (\ref{eq:BayesGGM}), but involves edge specific shrinkage parameters which are modeled using anatomical knowledge. Throughout this article, we will denote the structural connectivity metric as $p_{jk}$ for edge $(j,k)$, where a larger value denotes a stronger anatomical connection and vice-versa. For example, in our data application, $p_{jk}$ corresponds to the probability of SC obtained via probabilistic tractography. The proposed approach to estimating the brain functional network incorporating anatomical knowledge is defined as follows
\begin{eqnarray}
	\small \pi(\boldsymbol \Omega \mid \bfl) &=& C^{-1}_{\bfl,\nu}\prod_{k=1}^p Exp(\omega_{kk}; \frac{\nu}{2}) \prod_{j<k} DE( \omega_{jk}; \nu \lambda_{jk}) I(\boldsymbol \Omega\in M_p^{+}), \nonumber \\
	\pi(\bfl) &\propto& C_{\bfl, \nu} \prod_{j<k} LN(\mu_{jk} - \eta p_{jk}, \sigma^2_\lambda), \mbox{ } \mu_{jk}\sim \pi(\mu), \eta \sim \pi(\eta)1(\eta>0),\label{eq:siGGM}
\end{eqnarray}
where $\pi(\cdot)$ denotes the prior distribution, $\bfl = \{ \lambda_{jk}, j<k, j,k =1,\ldots,p\}$ denotes the collection of edge specific shrinkage parameters having a log-normal ($LN$) type distribution, $\nu$ refers to the tuning parameter controlling the network's overall sparsity and also corresponds to the scale parameter for the exponential prior on the diagonals, $\eta$ is a positive random variable which controls the average effect of SC on FC, $\mu_{jk}$ denotes the random edge specific baseline component representative of non-anatomical sources of variations regulating FC, and $C_{\bfl,\nu}$ is the intractable normalizing constant for the prior on the precision matrix depending on $\bfl$ and $\nu$. As in Wang et. al (2012), it can be shown that $C_{\bfl,\nu}$ is finite and cancels out the normalizing constant in $\pi(\bfl)$ when computing the posterior distribution $\pi(\Omega,\bfl\mid -)$, resulting in a closed form expression which facilitates computation. We note that in the extreme case when $\lambda_{jk}=\mu_{jk} - \eta p_{jk}$, the model specifies a parametric relationship for fixed choices of $\mu_{jk}$ and $\eta$, which has similarities with \cite{Ng2012} and \cite{Pineda2014}. 

\begin{figure}[h!]
 \centering
\mbox{\includegraphics[height=3in,width=1\textwidth,keepaspectratio]{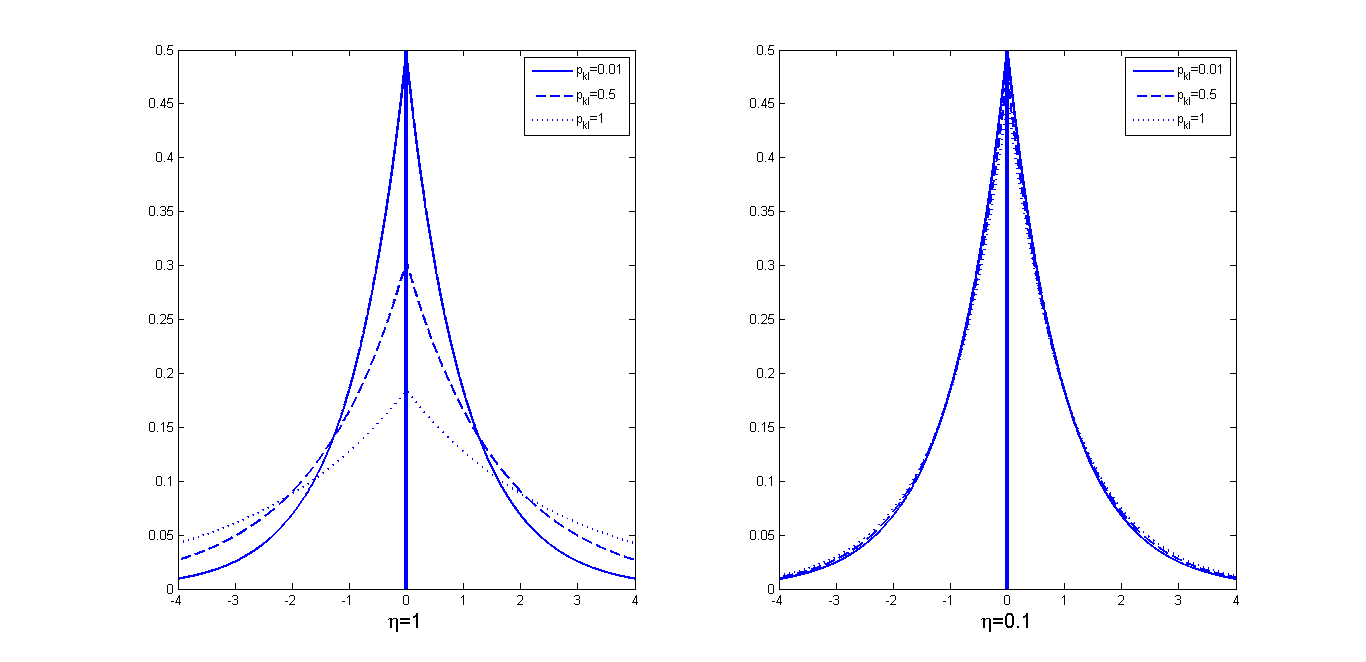}}
 \caption{\small Prior distribution of $\omega$ when $\lambda_{jk}=\mu_{jk} - \eta p_{jk}$.  $\mu_{jk}$ is fixed at zero with varying values of $\eta$ and SC. Solid, dashed and dotted lines correspond to $p_{jk}=$0.01,0.5, and 1, respectively. Left panel (a): for large $\eta$ values ($\eta=1$), the prior places increasing mass at the tails, which encourage stronger functional connectivity; right panel (b): for small values ($\eta=0.1$), the prior on $\omega$ does not change noticeably with the change in SC information.}\label{fig:omegavariability}
\end{figure}

The anatomically informed prior on the shrinkage parameters in (\ref{eq:siGGM}) specifies a probabilistic relationship between the edge specific shrinkage parameters and the given SC knowledge via $\eta$. In particular, increasing positive values of $\eta$ implies an increasing dependence on the given SC, potentially resulting in a functional connection even for small SC weights.  Figure \ref{fig:omegavariability} illustrates that for large $\eta$ and increasing SC, the Laplace prior has heavier tails and less mass around zero, which i interpreted as increased probability of strong FC.  In contrast, small values of $\eta$ do not result in a noticeable change in the prior distribution under varying SC strengths, implying a negligible relationship between SC and FC. Moreover, the shrinkage parameters are stochastically monotonically decreasing with respect to the SC strength, under the restriction $\eta>0$. This implies that as the SC strength ($p_{jk}$) for the edge $(j,k)$ is increased, the corresponding shrinkage parameter $\lambda_{jk}$ will take smaller values in probability, resulting in values of $\omega_{jk}$ which are away from zero. This encourages the presence of FC at edge $(j,k)$ for large values of $p_{jk}$. Similarly small values of $p_{jk}$ will encourage greater shrinkage for $\omega_{jk}$ resulting in the absence of FC at edge $(j,k)$.

Additionally, the baseline effect, $\mu$, corresponds to variations in underlying neuronal activity that are independent of the brain anatomy.  This formulation enables (a) more flexibility in the FC-SC relationship by allowing the possibility of strong FC when an anatomical connection is not obvious, and vice-versa; and (b) heterogeneity in FC across edges which possesses similar SC strength that is often encountered in practice. Overall, increasing(decreasing) absolute values of the baseline effect discourages(encourages) the presence of an edge in a manner that is independent of the anatomical information. The hyperparameters $\bfmu$ and $\eta$ are unknown and are learnt in a data-adaptive manner under the following priors
\begin{eqnarray}
\small \mu_{jk} \sim N(\mu_0,\sigma^2_\mu), j<k, j,k=1,\ldots,p,\mbox{ } \eta \sim Ga(a_\eta,b_\eta), \label{eq:hyper}
\end{eqnarray}
where $(\mu_0,\sigma^2_\mu)$, and $(a_\eta, b_\eta)$, are typically pre-specified.  The choices of these hyperparameters are described in Appendix \ref{eq:appendix}. We note that the scale parameter $\nu$ controls the overall network sparsity and is treated as a tuning parameter as in the penalized approach in (\ref{eqn:baseGGM}), enabling the estimation of a series of networks with varying sparsity levels. The optimal network is chosen as the point estimate corresponding to the value of $\nu$ minimizing the Bayesian Information Criteria.

\begin{figure}
\centering
\mbox{\includegraphics[height=3in,width=1\textwidth,keepaspectratio]{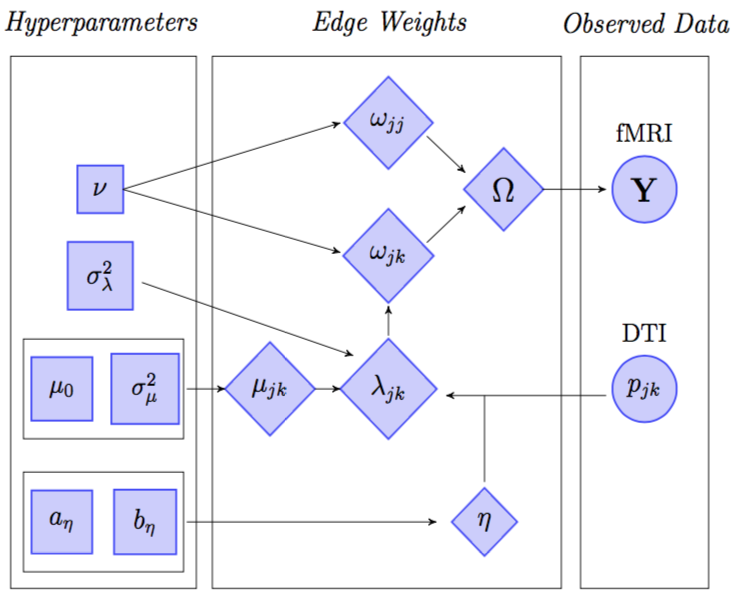}}
\caption{Graphical illustration of model parameters and their contribution to estimation of anatomically informed functional connectivity based on resting-state fMRI data \textbf{Y}.  Circles represent observed data, diamonds represent parameters to be updated, and squares represent fixed values.  }\label{fig:methoddiagram}
\end{figure}

\pagebreak
\subsection{Model Estimation}
Although the proposed model can be implemented using MCMC, it is not scalable to high dimensional settings involving a large number of nodes.  Moreover, MCMC samples require a post hoc thresholding step to select important edges since estimates cannot take exact zeros under a Laplace prior.  We bypass these limitations by computing a MAP estimate for the parameters of interest. Our iterative optimization approach employs an existing graphical lasso algorithm to sample the precision matrix given all other parameters, coupled with additional optimization steps to sample the shrinkage parameters and associated hyperparameters inherent in the Bayesian specification (\ref{eq:siGGM}). In order to fit the proposed model, we estimate $\Theta=(\boldsymbol{\Omega},\boldsymbol{\alpha},\eta,\boldsymbol{\mu})$ by maximizing the log-posterior distribution in (\ref{logpost}), where $\bfa=\log(\bfl)=\{ \log(\lambda_{jk}), j<k\}$ and $\boldsymbol{\mu}=\{ \mu_{jk}, j<k\}$ denotes the vector of edge-specific log-shrinkage parameters and baseline effects in (\ref{eq:siGGM}), respectively. Note that the posterior distribution can be written as $\small f(\Theta| {\bf y}_1,\ldots,{\bf y}_T) \propto f({\bf y}_1,\ldots,{\bf y}_T, \Theta) $
\begin{eqnarray}
\small
 & = & \prod_{t=1}^T N({\bf y}_t\mid 0, \boldsymbol{\Sigma}) \prod_{k=1}^p Exp(\omega_{kk}\mid \frac{\nu}{2}) \big\{\prod_{j<k} DE(\omega_{jk} \mid \nu,\lambda_{jk}) N(\alpha_{jk}\mid \mu_{jk},\eta, \sigma_{\lambda}^2) N(\mu_{jk}\mid \mu_0,\sigma_{\mu}^2)\big\} \nonumber \\
 & & \times Ga(\eta\mid a_\eta,b_\eta). \nonumber
\end{eqnarray}

We find the MAP solution for the model parameters by maximizing over the the posterior log-likelihood as $\small {\hat \Theta} =\underset {\Theta} {\arg\max   } \mbox{ } \dot{l}(\Theta )$, where
\begin{eqnarray}
\small
 \dot{l}(\Theta ) &=& -\frac{T}{2} log|\boldsymbol{\Omega}| + \frac{1}{2}tr(S|\boldsymbol{\Omega}|) +  \nu\sum_{j<k} e^{\alpha_{jk}}\lvert \omega_{jk} \rvert +\sum_{j<k}  \frac{(\alpha_{jk}-(\mu_{jk} - \eta p_{jk}))^2}{2\sigma_{\lambda}^2}  \nonumber \\
& -& (a_\eta-1)\log(\eta) + b_\eta \eta + \sum_{j<k}\frac{(\mu_{jk} -\mu_0)^2}{2\sigma_{\mu}^2} -p\log(\frac{1}{2}\nu) + \frac{1}{2}\nu\sum_{k=1}^p\omega_{kk}. \label{logpost}
\end{eqnarray}

All parameters in the posterior distribution are updated iteratively until convergence. The precision matrix is updated given other parameters using the existing graphical lasso algorithm, whereas $\mu_{jk}$ and $\eta$ are updated via closed form expressions and $\bfa$ is updated via a Newton-Raphson step since a closed form solution does not exist. The iterative updates continue until $\small \lvert \dot l(\Theta^{(m)})-\dot l(\Theta^{(m+1)})   \rvert < \epsilon \lvert \dot l(\Theta^{(m+1)}) \rvert$ for $\epsilon=10^{-4}$.  At convergence, $\small \hat \Theta = (\boldsymbol{\hat \Omega}, \hat \bfa, \hat \eta, \hat \bfmu)$ is the solution, where $\boldsymbol{\hat \Omega}$ is the anatomically informed functional brain network based on single subject data. We note that one could alternatively treat ${\bfmu}$ and $\eta$ as tuning parameters and compute a range of networks over a grid of $(\bfmu,\eta)$ values, and then select the optimal network as one minimizing some goodness of fit criteria. However, this strategy did not result in adequate numerical performance, highlighting the advantages of specifying suitable priors on hyperparameters in order to estimate them in a data-adaptive manner. The computational steps for updating the model parameters are detailed in Appendix A.

\section{Results and Discussion}

\subsection{Simulations}
\subsubsection{Simulation Setting} 	
 \noindent We conduct numerical studies to assess the performance of siGGM relative to SC naive and SC informed competitors.  SC naive approaches are representative of methods that do not incorporate auxiliary information, and includes the graphical lasso or \textit{Glasso} \citep{Friedman2007}, the partial correlation approach \textit{Space} proposed by \cite{Peng2009}, and the proposed approach in (\ref{eq:siGGM}) without structural information obtained by setting $\eta=0$, denoted \textit{siGGM}($\eta=0$). SC informed approaches incorporate anatomical information into the estimation routine.  We consider the Bayesian G-Wishart approach by \cite{Hinne2014} which treats FC as completely determined by SC and is denoted by G-Wishart and the adaptive graphical lasso approach by \cite{Pineda2014} which specifies a parametric relationship between the shrinkage parameters and SC, and is denoted by aGlasso. All of the above approaches, except {\it Space} which estimates partial correlations, calculate sparse inverse precision matrices, where a zero off-diagonal entry implies the absence of an edge. The \textit{Glasso} and \textit{Space} approaches are implemented via the R packages {\it glasso} and {\it space}, respectively.  We estimate the precision matrix under the G-Wishart approach using Matlab scripts available on the author's website and incorporate adaptive weights in the {\it glasso} algorithm to implement aGlasso. 

{\noindent \bf Data Generation:}  In order to assess the performance of our approach, we simulate data under three  assumed network structures and consider various relationships between SC and FC. The network structures are (a) Erdos Renyi (ER) networks consisting of edges randomly generated with probability $0.15$; (b) small-world (SM) networks generated under the Watts-Strogatz model \citep{Watts1998} in which most nodes may not be directly connected, but can reach other nodes via a small number of steps, and (c) scale-free (SF) networks generated using the preferential attachment model \citep{Barabasi1999}, in which nodes are more likely to link to a highly connected node than to a node with few connections, resulting in a hub network. For each network, we consider varying the number of nodes corresponding to $p=100, 200$. The data was generated using a Gaussian graphical model ${\bf y}_t \sim N_p(0, \boldsymbol{\Omega}^{-1}_\mathcal{G})$ with $T=200$ time points ($t=1,\dots,200$), where $\boldsymbol{\Omega}_\mathcal{G}$ is the precision matrix with zero off-diagonal elements corresponding to absent edges in the network $\mathcal{G}$.  Once $\mathcal{G}$ is determined based on the desired network structure, $\boldsymbol{\Omega}_{\mathcal{G}}$ is constructed as follows: the non-zero off-diagonal elements corresponding to important edges are generated from a Uniform(-1,1) distribution and the diagonal elements were fixed to be one. In order to ensure positive definiteness, we subtracted the minimum of the eigenvalues from each diagonal element of the generated precision matrix.  	

{\noindent \bf Prior SC Knowledge:} Conditional on $\boldsymbol{\Omega}_{\mathcal{G}}$, we construct several types of SC information, where the SC strengths were generated randomly between $(0,1)$ according to the following schemes. For scenario $MI$, we specify that 50\% of those edges with strong FC also have strong SC($>0.7$), while 25\% of those edges with strong FC have moderate SC($0.3-0.7$), and the remaining 25\% have weak SC($0-0.3$). For scenario $MII$, the proportion of edges that have strong FC, coupled with strong SC, moderate SC, and weak SC, are 30\%, 35\% and 35\% respectively. For each of the scenarios $MI$ and $MII$, we also consider two levels of mis-specification of the SC information, which are denoted as $MI(a), MI(b),$ and $MII(a),MII(b),$ respectively. For $MI(a)/MII(a)$, we specify that 10\% of those edges with zero FC have non-zero SC, while for $MI(b)/MII(b),$ we fix 20\% of those edges with zero FC to have non-zero SC. All the other edges with zero or weak FC are assumed to have small SC, while remaining edges with moderate FC have non-zero SC strengths. We note that edges having zero FC but non-zero SC represent potential mis-specification of anatomical knowledge, based on the notion that strong SC typically underlies robust non-zero FC \citep{Shen2015, Kemmer2017}. 

{\noindent \bf Comparison Metrics:} To assess the performance under different approaches, we compute the area under the curve (AUC), which is a measure of the estimated sensitivity versus specificity over different network sparsity levels. Sensitivity is computed as $TP/(TP + FN)$, while specificity is defined as $TN/(TN + FP)$, where $TP, TN, FP,FN$ denote the number of edges that are true positives, true negatives, false positives and false negatives, respectively. To evaluate the point estimate of the network obtained under the Bayesian information criteria (BIC), we compute the Matthew's correlation coefficient (MCC) which is a scalar measure combining sensitivity and specificity and is defined as MCC$=\small \frac{TP\times TN-FP\times FN}{\sqrt{(TP+FP)(TP+FN)(TN+FP)(TN+FN)}}$ \citep{Wang2012}. We compute the relative $L_1$ norm error, $\small(| \hat \Omega - \Omega|_1)/|\Omega|_1$, to assess accuracy in estimating the precision matrix encapsulating the FC strengths. Since brain networks are also often evaluated in terms of summary statistics reflecting network organization, we evaluate the accuracy in estimating the global efficiency which is a commonly used measure for global integration of brain connectivity.

\subsubsection{Results}
The results under siGGM and SC naive approaches under $MI(a)$ are presented in Table \ref{table:vsScmethods},  and Table \ref{table:misspecify} displays results for SC informed approaches under various levels of mis-specification.  Table \ref{table:vsScmethods} illustrates that either the proposed siGGM approach, or the variant of the proposed approach with no prior knowledge ($\eta=0$), have the lowest bias in estimating the global efficiency for all dimensions. Moreover, the proposed approach always has a higher MCC and AUC values, and a lower $L_1$ error norm compared to all other SC naive approaches. These results demonstrate the advantages of using structural knowledge to guide network estimation. 

When the mis-specification levels are varied, Table \ref{table:misspecify} illustrates that the proposed method has a consistently lower bias in estimating the global efficiency for both $10\%$ and $20\%$ mis-specification levels, compared to alternative SC informed approaches. Moreover, while the G-Wishart approach may have a higher MCC for $p=100$ when the mis-specification level is 10\% (cases $MI(a)$ and $MII(a)$), the proposed method has a comparable or higher MCC for 20\% mis-specification levels (cases $MI(b)$ and $MII(b)$). Moreover under $p=200$, the MCC under the proposed approach is the highest for small-world and scale-free networks, and comparable to the G-Wishart method for the Erdos-Renyi network. We also note that while aGlasso often has the lowest MCC values, it may sometimes yield a higher AUC under small world and scale-free networks for $p=100$. However, the proposed approach is shown to have the highest AUC for $p=200$ for all scenarios, highlighting the advantages of incorporating prior knowledge in a flexible manner in higher dimensions. Finally, siGGM consistently has the lowest $L_1$ error in estimating the precision matrix across all networks and dimensions. The above results illustrate a robust performance of the proposed method for $p=100$ and a superior performance for $p=200$ under the small world and scale-free networks, which closely resemble brain networks encountered in practical applications.

\begin{figure}[h!]
 \centering
\mbox{\includegraphics[height=3in,width=1\textwidth,keepaspectratio]{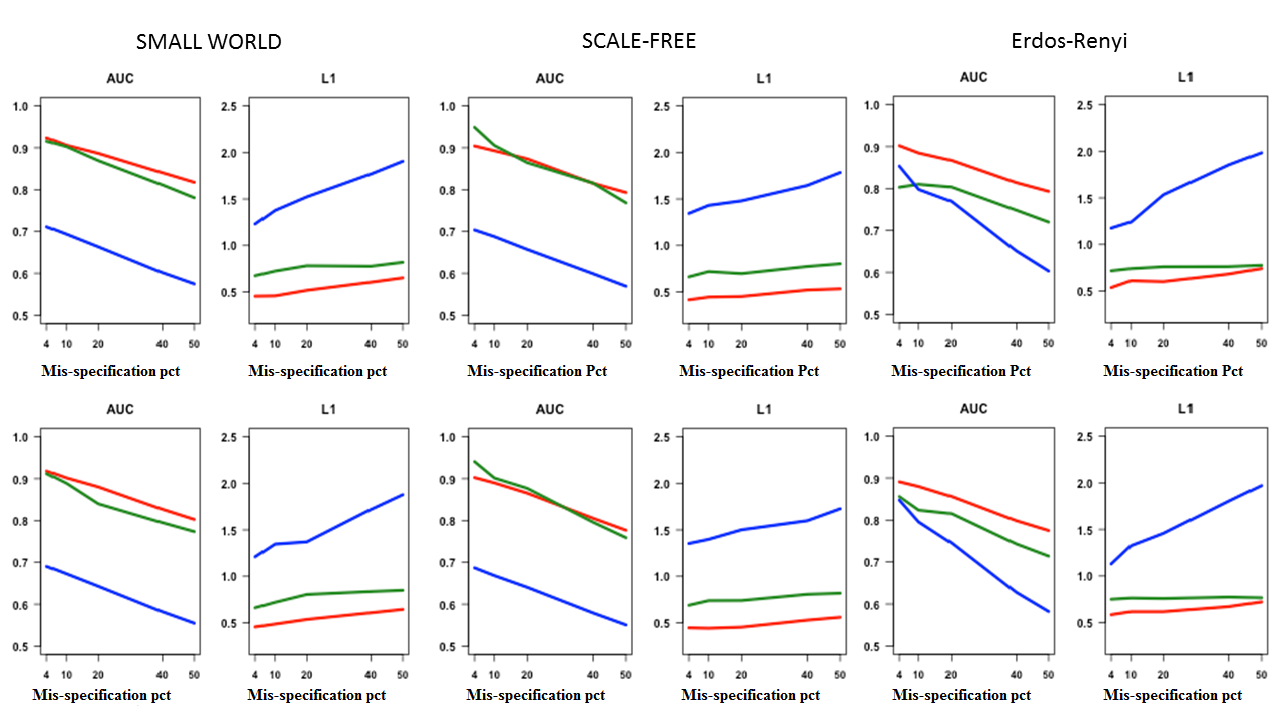}}
 \caption{\small Comparison of siGGM (\textbf{red}), G-Wishart (\textbf{blue}), and aGlasso (\textbf{green}) simulation results for different network structures with p=100 regions under scenario MI (top row) and MII (bottom row).  Each panel displays the AUC or L1 relative error as the percentage of conditionally independent edges with non-zero anatomical connectivity increase.}  
\label{fig:simmetrics}
\end{figure}

Although Table \ref{table:misspecify} provides some idea about the relative performance under mis-specification, it is of interest to look at the effects of mis-specification in more detail. Hence, we examined the AUC and $L_1$ error values as the mis-specification level was gradually increased from 4\% to 50\%, under different networks for $p=100$. The results, presented in Figure \ref{fig:simmetrics} illustrate that the proposed method consistently has higher AUC under the Erdos-Renyi network, and has a larger AUC for non-trivial mis-specification levels under the small world network. Further, the gains in AUC under the proposed approach seems to steadily increase with growing mis-specification levels under both these networks. Under the scale-free network, aGlasso has higher AUC under lower mis-specification levels when $p=100$, but the proposed approach exhibits comparable or higher AUC as the mis-specification level is increased and it has a higher AUC for larger dimensions ($p=200$) as in Table \ref{table:misspecify}. For all networks, the proposed method is seen to have the lowest $L_1$ error across all mis-specification levels, while the error under the G-Wishart increases sharply as the mis-specification level is increased. These results provide clear evidence that while alternative SC informed approaches may occasionally perform better for lower mis-specification levels and moderate dimensions, the proposed siGGM method demonstrates a superior performance when the mis-specification level as well as the number of nodes is increased, which is of paramount importance in whole brain connectome studies involving high resolution atlases \citep{Glasser2016, Power2011} and non-trivial mis-specification of the anatomical information.

 Finally, we note that the siGGM can be implemented fairly quickly.  On a 2.5Gz Intel Core i5 processor, the procedure estimates the optimal graph structure in approximately three seconds for p=40,  twenty seconds for p=100, and approximately four minutes for p=200.   While these computation times are slightly slower compared to generic graphical modeling approaches naive to anatomical knowledge, the overall computation is sufficiently quick and feasible for practical implementation in whole brain connectome analysis.

\subsection{PNC Data Application}\label{sect:results}
Existing literature has examined various neural substrates for age-related changes using structural and functional neuroimaging \citep{Gur2012, Shaw2008, Raznahan2011}. Moreover, gender differences have been extensively documented in behavioral measures \citep{Halpern2007, Hines2010}, structural neuroimaging \citep{Lenroot2007}, and functional imaging measures \citep{Lenroot2010}. However, gender related differences in the developmental trajectory of the brain functional network from childhood to adolescence are still not understood well \citep{Gur2012}, and further, limited attempts have been made to investigate such differences by fusing functional and structural neuroimaging data. We use resting-state fMRI and DTI data from the Philadelphia Neurodevelopment Cohort (PNC) study to obtain preliminary answers to these questions.  After estimating brain functional connectivity based on SC knowledge, we examine FC differences  between boys and girls across different age groups. 

We perform the analysis separately for each gender within the three age groups 8-12 (pre-teen), 13-17 (teen), 18-21 (young adult), where each age group contains approximately 9 to 12 individuals, and is constructed similarly to those in \cite{Ingalhalikar2014}.  All subjects are right-handed, physically, and mentally healthy, enabling a fair comparison between the groups. In addition to assessing gender based network differences, we also perform a secondary analysis to assess our method's ability to reliably estimate functional networks. For this analysis, we split each subject's resting-state fMRI time series into two equally sized scanning sessions (60 scans each) and calculate the intraclass correlation coefficient (refer to equation (\ref{eq:ICC})) for seven network metrics which are widely used to summarize brain networks. These metrics include clustering coefficient, characteristic path length, local efficiency, global efficiency, modularity, hierarchy, and degree, and they were calculated with the {\it Matlab} toolboxes Brain Connectivity Toolbox \citep{Rubinov2010} and GRETNA \citep{wang2015gretna}. 

\subsubsection{Data preprocessing}
Resting-state fMRI scans were acquired on a single-shot, interleaved multi-slice, gradient-echo, echo planar imaging (GE-EPI) sequence \citep{Satterthwaite1}.  Nominal voxel size is 3x3x3mm with full brain coverage achieved with the following parameters: TR/TE=3000/32 ms, flip=90$^\circ$, FOV=200 $\times$ 220 mm, matrix= 64 $\times$ 64, 46 slices, slice thickness/gap=3 mm/0 mm for a total of 6.2 minutes.  Participants were instructed to remain awake, motionless, and fixated on a crosshair throughout the duration of the data acquisition.   Several standard preprocessing steps were applied to the rs-fMRI data, including despiking, slice timing correction, motion correction, registration to MNI 2mm standard space, normalization to percent signal change, removal of linear trend, regressing out CSF, WM, and 6 movement parameters, bandpass filtering (0.009 to 0.08), and spatial smoothing with a 6mm FWHM Gaussian kernel.  Subsequent voxel level data is aggregated into 90 regions of interest (ROI) based on the Automated Anatomical Labelling atlas \citep{Tzourio2002}.  For each ROI, the average time series of all constituent voxels represents the region's temporal BOLD signal.

Diffusion weighted images permit us to localize and orient white matter fiber bundles via the diffusion of water in the brain.  Images were acquired on a twice-refocused spin-echo (TRSE) single-shot EPI sequence for a total of 64 diffusion-weighted directions with b=1000 s/mm$^2$ and 7 scans with b=0 s/mm$^2$ \citep{Satterthwaite1}.  Acquisition parameters were TR/TE=8100/82ms, matrix=128$\times$128, FOV=240mm, slice thickness=2mm, GRAPPA factor=3.  Due to gradient induced vibrations disturbing image quality, DWI images were acquired in two imaging runs to reduce the continuous duration in which subjects tolerate the scan.  Standard pre-processing procedures, such as eddy current correction and bias-field correction are applied to the diffusion weighted data.  Subsequently, we use the FSL functions bedpostx and probtracx2 to estimate the distribution of fiber tensors at each voxel and the count of white matter fibers tracts connecting all pairs of brain regions, respectively.

\subsubsection{Results}
\begin{figure}[h!]
 \centering
\begin{adjustbox}{minipage=\linewidth,scale=0.65}
\begin{subfigure}{.5\textwidth}
  \centering
  \includegraphics[width=.8\linewidth]{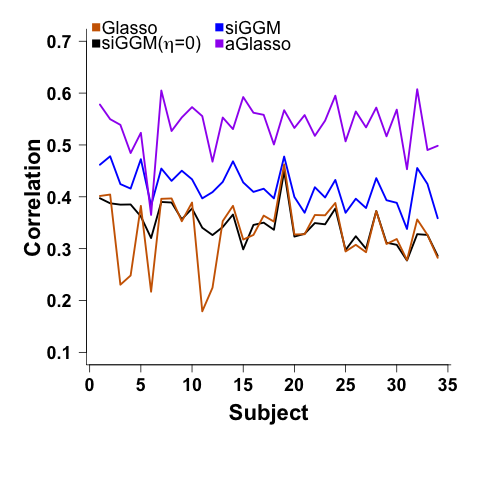}
  \caption{}
\end{subfigure}%
\begin{subfigure}{.5\textwidth}
  \centering
  \includegraphics[width=.8\linewidth]{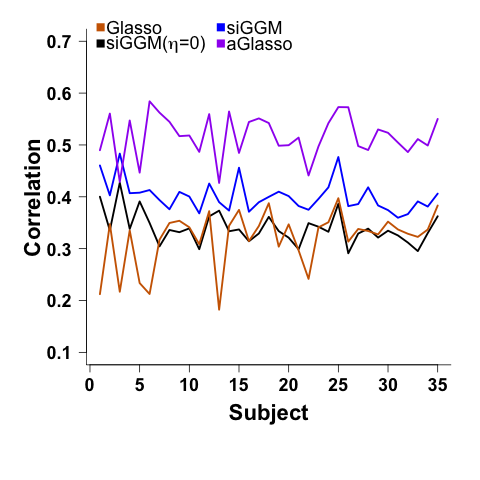}
  \caption{}
\end{subfigure}
\begin{subfigure}{.5\textwidth}
  \centering
  \includegraphics[width=.8\linewidth]{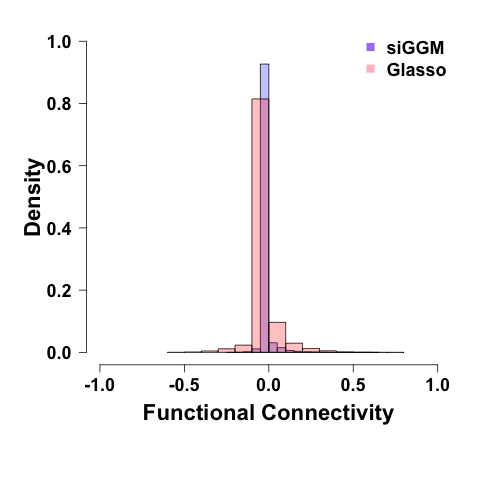}
  \caption{}
\end{subfigure}%
\begin{subfigure}{.5\textwidth}
  \centering
  \includegraphics[width=.8\linewidth]{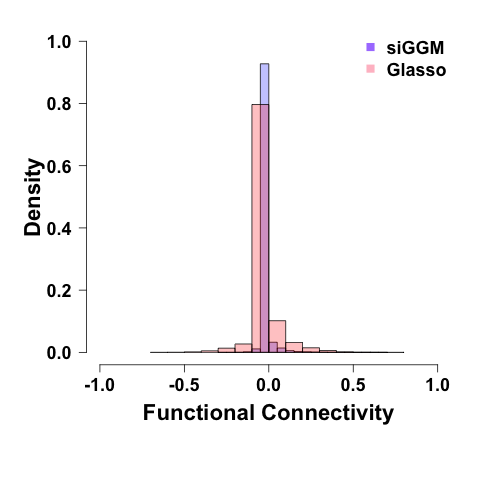}
  \caption{}
\end{subfigure}
\end{adjustbox}
 \caption{\small Assessment of associations between anatomical and functional connectivity strength for all males ({\it left column}) and females ({\it right column}) in the study.  (A) and (B) display the correlation between estimated functional connectivity and non-zero anatomical connectivity weights for {\it glasso}, aGlasso, \textit{siGGM}($\eta=0$), and siGGM.  (C) and (D) depict the distribution of functional connectivity for regions not directly connected. Methods incorporating anatomical information produce functional connections adhering to brain structure more closely than anatomically naive methods.}  \label{fig:correlations}
\end{figure}

In Figure \ref{fig:correlations} (A) and (B), we see that siGGM functional connectivity estimates are more highly correlated with non-zero anatomical connection strength across all subjects compared to Glasso and \textit{siGGM}($\eta=0$), but have a lower correlation compared to the aGlasso approach specifying a parametric structure-function relationship. Additionally, we note in panels (C) and (D) of Figure \ref{fig:correlations} that siGGM leads to larger shrinkage and smaller variance for conditional dependencies between anatomically isolated brain regions compared to the generic graphical lasso without prior knowledge.  This yields a smaller number of functional connections between anatomically disconnected ROIs. The above results indicate that our method adheres more closely to the brain's anatomical connectivity without restricting the FC to be a deterministic function of structural pathways.

\begin{figure}[h!]
 \centering
 \begin{adjustbox}{minipage=\linewidth,scale=0.65}
\begin{subfigure}{.75\textwidth}
  \centering
  \includegraphics[width=.97\linewidth,height=.47\textheight]{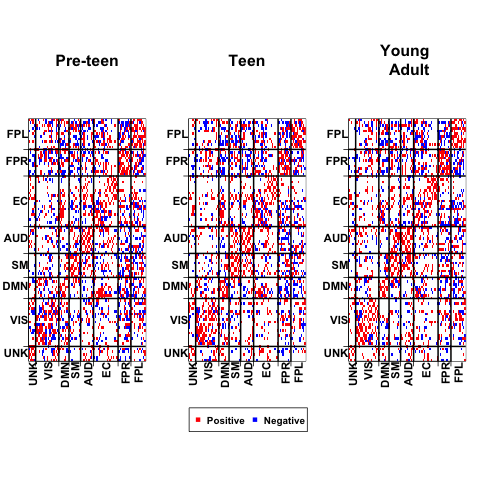}
  \caption{Female}
\end{subfigure}%

\begin{subfigure}{.75\textwidth}
  \centering
   \includegraphics[width=.97\linewidth,height=.47\textheight]{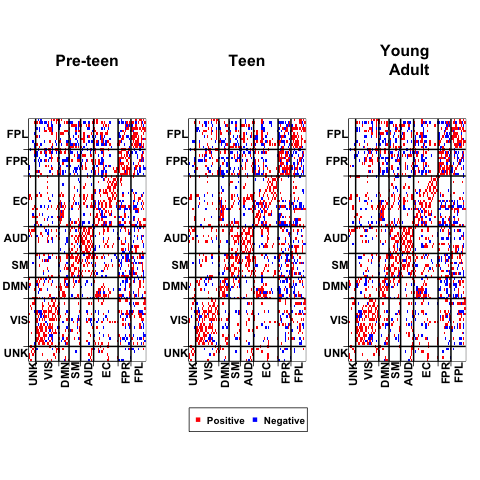}
   \caption{Male}
\end{subfigure}%
\end{adjustbox}
 \caption{\small Average network estimates for females ({\it top row}) and males ({\it bottom row}) in each of the age ranges.  Reported edges are those with partial correlation value exceeding .005 in magnitude.  While both sexes have similar network structures across the three age groups, female networks exhibit slightly increased connectivity relative to the networks of males.  }\label{fig:averagenetworks}
\end{figure}

\begin{figure}[h!]
 \centering
\begin{adjustbox}{width=1\linewidth}
\begin{subfigure}{.65\textwidth}
  \centering
  \includegraphics[width=.87\linewidth]{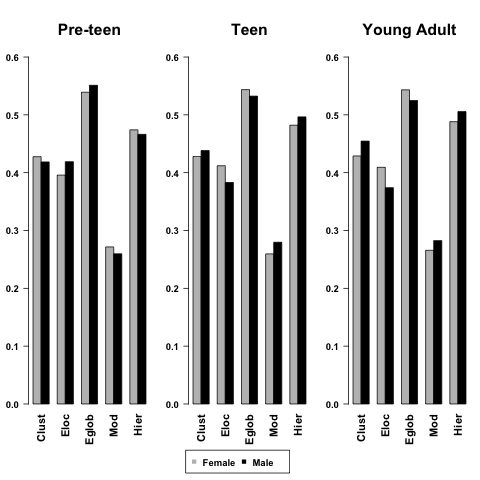}
  \caption{}
\end{subfigure}%

\begin{subfigure}{.65\textwidth}
  \centering
   \includegraphics[width=.87\linewidth]{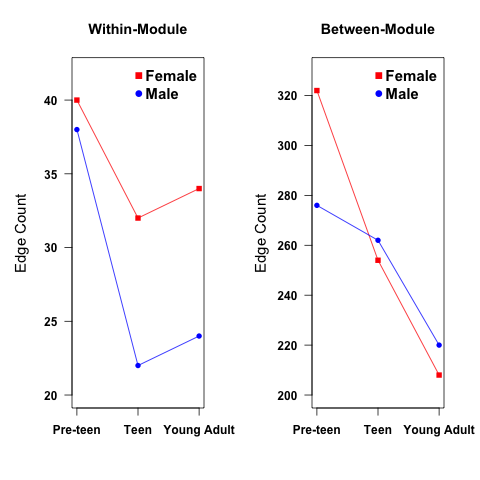}
  \caption{}
\end{subfigure}%
\end{adjustbox}
\caption{\small Topological features of estimated networks in males and females across the three age groups.  (A) displays network properties averaged over the respective gender and age group; (B) displays differentially weighted edges within- and between- module stratified by gender.  In teens and young adults, females have more within module connections and fewer between module connections than males.}\label{fig:betweenwithin}
\end{figure}

For network analysis, we classify each ROI into one of eight functional modules corresponding to resting state networks as defined in \cite{Smith2009}. These functional modules include a medial visual network, an occipital pole and lateral visual network (``VIS", 18 nodes), the default mode network (``DMN", 8 nodes), a sensorimotor network (``SM", 9 nodes), an auditory network (``AUD", 10 nodes), an executive control network (``EC", 19 nodes), right and left frontoparietal modules (``FPR" and ``FPL", 11 and 10 nodes, respectively) and an unknown module containing unassigned nodes (``UNK", 5 nodes). Figure \ref{fig:averagenetworks} shows that males and females have similar connectivity patterns with primarily positive connections within functional modules.  Further comparisons of male and female brain networks within each age group reveals that consistent connections across age groups persist within module while inconsistent connections mainly exist between modules.  After standardizing by the number of nodes in each module, the SM and AUD were found to be the two most highly connected functional modules in males and females across all age groups.  Figure \ref{fig:betweenwithin} (A) illustrates the similarity in network architecture for males and females with all metrics having non-significant differences across genders (except for local efficiency for teens and young adults), which implies shared patterns in brain organization across gender and age groups. Figure \ref{fig:betweenwithin} (B) illustrates  that  males exhibit greater between module but smaller within-module connectivity differences in teens and young adults, which is supported by previous work and has been linked to variations in emotional identification and spatial cognitive tasks \citep{Satterthwaite2}. However, we discover  greater between- and within-module connections in pre-teen girls, which is an interesting finding that requires further examination. 

As a second level of the analysis, we are also interested in the distribution of differentially weighted edges (DWE) between males and females, within each of the eight functional modules. DWE were identified as connections for which the FC strength was significantly different between genders under a permutation test. To evaluate if the number of DWEs within and between modules occur more often than allowed by chance, we define a goodness of fit measure (equation (\ref{chi}) in Appendix A) which represents the deviation between observed and expected numbers of DWE for each module block, standardized by the expected number. This measure captures whether a given module block has unusually high or low occurrence of DWE and enables us to identify modules with the most pronounced differences across gender. From the results  presented in Table \ref{table:permwithin}, we discover statistically significant differences in the number of DWEs occurring in the executive control (EC) module in pre-teens and young adults, which is supported by previous results on gender related differences in the EC \citep{Hyde1981, Mansouri2016}. Table \ref{table:permwithin} also suggests that gender based differences attenuate with development, with the largest number of DWEs in the pre-teen group and the smallest in the young adult group. We also find the DWE between the cingulum\_ant\_L in the EC and parietal\_inf\_L in the DMN exists in pre-teens, teens, and young adults, which suggests consistent gender based differences during the developmental phase. These regions are known to have brain volume differences between males and females which may point to subtle cognitive variations \citep{Frederikse1999, Ruigrok2014}.

\begin{figure}[h!]
 \centering
 \mbox{\includegraphics[height=3in,width=.5\textwidth]{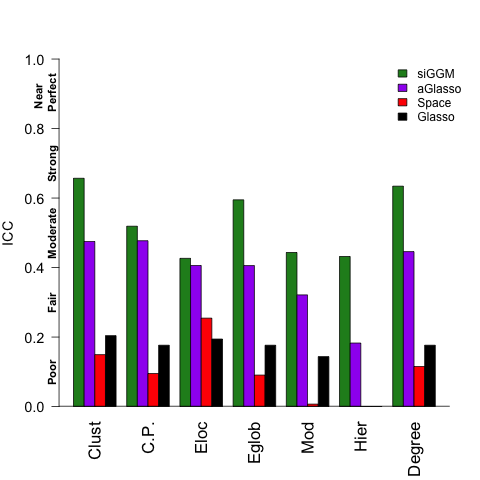}}
   \caption{\small Reliability of network metrics across scanning sessions for siGGM, aGlasso, {\it Space}, and {\it glasso}.  We estimate seven network attributes (clustering coefficient (Clus.), characteristic path length (C.P.), local efficiency (Eloc), global efficiency (Eglob), modularity (Mod), hierarchy (Hier), and degree (Degree)) and report ICC(3,1) for all subjects.  ICC values are classified according to the agreement scale $0<ICC\le .2$ (poor), $.2<ICC\le .4$ (fair), $.4<ICC \le .6$ (moderate), $.6<ICC\le 8$ (strong), and $.8<ICC\le 1$ (near perfect) as suggested by Telesford et. al (2010).  Anatomically informed FC estimates produce more reliable networks than SC naive methods.}  \label{fig:metricsreliability}
\end{figure}

A major challenge in resting state connectivity studies is to ensure reproducibility of the findings \citep{Griffanti2016}. We demonstrate that appropriately incorporating anatomical connectivity information leads to stable topological features of estimated networks across scanning sessions.  Figure \ref{fig:metricsreliability} displays the intraclass correlation coefficient (ICC) of seven network metrics under different approaches, where the details for computing the ICC are outlined in equation (\ref{eq:ICC}) in Appendix A. It is clear that the proposed siGGM produces estimates that have notably larger ICC measures for all the network metrics compared to all the other approaches considered. The reproducibility under the proposed approach is substantial for the clustering coefficient, global efficiency, and degree, and is moderate for all the other metrics. Moreover, it is reassuring to see that these three metrics with the highest ICC values under the proposed approach have been shown to be the most reproducible network metrics in independent studies \citep{Niu2013, Telesford2010, Wang2011graph}.  In contrast, reproducibility is barely moderate under aGlasso for most metrics and weak under SC naive approaches.  These findings highlight the benefits of incorporating anatomical information in a flexible manner. Although not presented, we note that the G-Wishart approach leads to an unrealistic ICC value of one in all cases, which is starkly different than the reliability values reported in previous studies  \citep{Welton2015}. The perfect reliability is due to the fact that G-Wishart relies entirely on the SC information for specifying the functional network structure, resulting in the exact same network for both the sessions. Hence the reproducibility results under G-Wishart are not comparable. 

\clearpage

\section{Conclusion}

In this paper, we introduce a novel Bayesian approach and an associated optimization algorithm for fusing structural and functional imaging data in order to estimate brain functional networks. We propose a flexible method for incorporating {\it apriori} known anatomical connectivity information in order to estimate the functional brain networks, bypassing the limitations of existing approaches by accommodating complex structure-function relationships while allowing unknown sources of variation independent of underlying anatomical structure. The proposed model is biologically more realistic compared to existing methods and often outperforms alternative approaches with or without anatomical knowledge as illustrated via extensive numerical studies. In particular, the advantages under the proposed method become more evident as the mis-specification levels for the anatomical knowledge and/or the number of nodes is increased, which has important practical implications. An analysis of the PNC data yields brain networks which have strongly reliable network metrics under the proposed approach, whereas the reproducibility under other approaches are moderate at best.  

While we primarily focus on direct anatomical connections between brain regions, siGGM readily incorporates complex measures of anatomical connectivity corresponding to more indirect connections.  In future work, we intend to explore various direct and indirect measures of structural connectivity so as to assess which anatomical measures yield the most meaningful and reproducible FC results.



\begin{appendices}
\section{}\label{eq:appendix}
\setcounter{equation}{0}
\renewcommand{\theequation}{\thesection.\arabic{equation}}

\subsection{Parameter Updates}
We optimize (\ref{logpost}) by iteratively updating model parameters as follows.

{\noindent \textbf{Update} $\boldsymbol \Omega$}: Given the data, ${\bf Y}=\{{\bf y}_1,...,{\bf y}_T\}$, and current estimates for all other model parameters, we solve 
\begin{equation}\label{Omegaupdate}
\begin{split}
\boldsymbol{\hat \Omega^{(m+1)}} &=\underset {\boldsymbol \Omega} {\arg\min }  -\log\lvert \boldsymbol \Omega \rvert + tr(S\boldsymbol \Omega)+ \frac{\nu}{2}\sum_{j < k}  e^{\alpha_{jk}^{(m+1)}}\lvert\omega_{jk} \rvert +  \frac{1}{2}\nu\sum_{j=k}\lvert\omega_{kk}\rvert \\
\end{split}
\end{equation}
for $ \boldsymbol{\hat \Omega}$.  This resembles the penalized likelihood framework of the traditional Gaussian graphical model.  Define $\delta_{ij}^{(k+1)}=\frac{1}{2}\exp(\alpha_{jk}^{(m+1)})$ for $j \ne k$ and $\delta_{jk}^{(m+1)}=\frac{1}{2}$ for $j=k$.  We can re-express (\ref{Omegaupdate}) as
\begin{equation*}
\begin{split}
\boldsymbol{\hat \Omega^{(m+1)}} &=\underset {\boldsymbol \Omega} {\arg\min } -\log\lvert \boldsymbol \Omega \rvert + tr(S\boldsymbol \Omega)+ \nu\sum_{j< k} \delta_{jk}^{(m+1)}\lvert \omega_{jk} \rvert,
\end{split}
\end{equation*}
where we update $\boldsymbol \Omega$ using a quadratic approximation solver, \textit{QUIC}, available in R.   \\

{\noindent \textbf{Update  $\mu_{jk}$:}} Given {\bf Y}, $\eta^{(m+1)}$, and $\alpha_{jk}^{(m)}$, we update $\mu_{jk}$ via the closed form equation
$$\mu_{jk}^{(m+1)} = \frac{\sigma_{\mu}^2(\alpha_{jk}^{(m)}+\eta^{(m+1)}p_{jk})+\sigma_{\lambda}^2\mu_0}{\sigma_{\mu}^2+\sigma_{\lambda}^2}
$$.

{\noindent \textbf{Update  $\eta$:}} Given {\bf Y} and $\boldsymbol{\alpha}^{(m)}$, we can update $\eta$ via closed form equation
$$\hat \eta^{(m+1)} = \frac{-\beta^{(m)} + \sqrt{(\beta^{(m)})^2-4\gamma\rho}}{2\gamma},$$
where $\beta^{(m)}=b_\eta + \frac{\sum_{j<k} \alpha_{jk}^{(m)} p_{jk}}{\sigma_{\lambda}^2} - \frac{1}{\sigma_{\lambda}^2} \sum_{j<k}\mu_{jk}^{(m)}p_{jk} $, $\gamma=\frac{\sum_{j<k} p_{jk}^2}{\sigma_{\lambda}^2}$, and $\rho=-\frac{1}{\sigma_{\lambda}^2}(a_\eta-1)$

{\noindent \textbf{Update $\bfa$:}} Given {\bf Y}, $\boldsymbol{\Omega}^{(m)}$, $\mu^{(m+1)}$, and $\eta^{(m+1)}$, we can estimate $\alpha_{jk}^{(m+1)}$ for $1\le j < k\le p$ by solving
$$\hat {\bfa}^{(m+1)} = \underset {\bfa} {\arg\min } \quad \nu\sum_{j<k} e^{\alpha_{jk}}\lvert \omega_{jk}^{(m)} \rvert +\sum_{j<k}\frac{(\alpha_{jk}-(\mu_{jk}^{(m+1)}-\eta^{(m+1)} p_{jk}))^2}{2\sigma_{\lambda}}$$

A closed form solution doesn't exist, so we implement a Newton Raphson solver to find the optimal choice of $\bfa$.  Re-expressing this problem, we have
$$ \underset {\bfa } { \arg\min  } \quad exp( \bfa)' \lvert  \omega^{(m+1)} \rvert- \frac{1}{2\sigma_{\lambda}^2}(\bfa -( \mu_{jk}^{(m+1)}-\eta^{(m+1)} \tilde P))^{'}(\bfa -( \mu_{jk}^{(m+1)}-\eta^{(m+1)} \tilde P)) $$
where $\bfa=\{ \alpha_{12},\alpha_{13},...,\alpha_{(p-1)p}   \}$, $\tilde P$ denotes the upper diagonal elements of the structural connectivity matrix $P$, $e^{ \bfa}$ is the element wise exponential for each component of $\bfa$, and {\textbf 1} as a vector of 1's of length $\frac{p(p-1)}{2}$. Since $\boldsymbol \Omega$ is symmetric and we do not shrink diagonal elements, we simplify our estimation of $\bfa$ by only focusing upon the upper diagonal elements.

The Newton Raphson updating equation based on step size $\Delta$ is $\bfa^{m+1} = \bfa^{m}- \Delta g(\bfa^{m})H(\bfa^{m})^{-1},$
where $g(\bfa)=\nu \sigma_{\lambda}^{2} D_{\lvert \omega^{(m)} \rvert} e^{\bfa} +[\bfa -(\bfmu^{(m+1)}- \eta^{(m+1)} \tilde P)]$ and $H(\bfa)= \nu \sigma_{\lambda}^{2} D_{\lvert \omega^{(m)} \rvert} D_{\lvert e^{\bfa} \rvert} + I$,  $D_{\lvert \omega^{(m)} \rvert}$ is a $\frac{p(p-1)}{2} \times \frac{p(p-1)}{2}$ diagonal matrix with elements as the upper triangular elements of $\bfO$, and similarly for $D_{\lvert e^{\alpha} \rvert}$, and $I$ is an identity matrix.  Since $H$ is a diagonal matrix, it is easily inverted and serves as an appropriate Hessian matrix.  We search for the step size ($\Delta$) using a back tracking line search for each update of $\bfa$ as in Chang et. al (2017).

\subsection{Hyper-parameter Choice and Initial Values}
The proposed siGGM approach iteratively solves for the MAP estimator and works best when reasonable starting values are provided. We first find an initial estimate for the graph structure and the sparse inverse precision matrix ($\boldsymbol{\Omega}_0$), using the graphical lasso.  We initialize all edge specific penalty parameters as $\lambda_0$, which is the global tuning parameter corresponding to $\boldsymbol{\Omega}_0$. We set $\sigma^2_{\mu}=5,$ corresponding to an uninformative prior which  reflects our lack of knowledge regarding the baseline effects and choose $\mu_0=0$ as a default setting. We randomly generate the edge specific baseline effects $\mu_{jk}$ from the prior distribution $N(\mu_0,\sigma^2_\mu)$ and use these as initial values. The initial value of $\eta$ is chosen by averaging $-\frac{\sum_{l<k} (\exp(\lambda_0) - \mu_{jk})/p_{jk}}{p(p-1)/2}$, which is the average of all possible $\eta$ values under the relationship $\exp(\lambda_0) = \mu_{jk} - \eta p_{jk}, j<k$ corresponding to $\sigma_\lambda=0$. We choose $\sigma^2_\lambda = \frac{1}{p(p-1)/2} \sum_{j<k}\sum_{j,k=1}^p (\exp(\lambda_0) - \mu_{jk} - \eta p_{jk})^2 $.

Finally, we found that choosing $a_\eta$ and $b_\eta$ to attain E$[\eta]\approx 6$ and Var$[\eta]\approx1$ allows incorporation of structural information in a flexible manner.  However, larger first moments for the prior on $\eta$ may lead to increased false positives as our method places more weight on smaller structural connections, and similarly, smaller first moment may decrease the overall impact of structural information.  For example, when $a_\eta>1$ and $b_\eta\rightarrow \infty$, we have E$[\eta]\rightarrow 0$, which makes the siGGM  indistinguishable from SC naive methods.  

\subsection {Measure for computing between module differences}
  We define the goodness of fit measure
\begin{equation}
\label{chi}
X^{2}_{g_1,g_2}=\frac{(Q_{(g_1,g_2)}-E_{(g_1,g_2)})^{2}}{E_{(g_1,g_2)}},
\end{equation}
where $g_1,g_2 \in \{1,\ldots,G\}$ are the indices corresponding to one of the $G$ functional modules, $Q_{g_1,g_2}$ represents the observed number of DWEs in the $(g_1, g_2)$ block, $E_{g_1,g_2}$ represents the expected number of DWEs in the $(g_1, g_2)$ block when edges distribute randomly across the module blocks. $X^2_{g_1,g_2}$ measures the goodness of fit for each within-module block ($g_1=g_2$) or between-module block ($g_1\ne g_2$). In equation (\ref{chi}), the expected value can be derived in a straightforward manner as $E_{g_1,g_2}= 0.5p^{\ast}\{|g_1|(|g_2|-1)\}$ for within module blocks ($g_1=g_2$) and $E_{g_1,g_2}=p^{\ast}|g_1||g_2|$ for between-module blocks ($g_1 \ne g_2$), where $|g|$ represents the total number of nodes within the $g$th module, and $p^{\ast}$ represents the proportion of DWE among all the edges across the network. Using 5000 permutations of group labels at each edge, the DWE are identified as those connections with significant FDR-adjusted p-values.

\subsection{Calculation of ICC}
The intraclass correlation coefficient is a widely used reliability metric for assessing test-retest reliability of brain network topology in neuroimaging applications.  Using ICC(3,1), (two-way mixed single measures testing for consistency) we investigate the reliability of graph metrics across two scanning session \citep{Guo2012, Telesford2010}. The quantity is calculated as
\begin{equation}\label{eq:ICC}
ICC(3,1)=\frac{BMS-EMS}{BMS+(k-1)EMS},
\end{equation}
\noindent where $k$ is the number of scanning sessions per participant, $BMS$ is the between mean square and $EMS$ is the mean residual sum of squares.  BMS captures the variability between subjects while EMS measures unexplained within-subject variation in functional connectivity across scanning sessions (see \cite{Shrout1979}).  This metric is commonly used to measure test-retest network stability in brain networks (Braun et al., 2012) with agreement scale $0<ICC\le .2$ (slight), $.2<ICC\le .4$ (fair), $.4<ICC \le .6$ (moderate), $.6<ICC\le 8$ (strong), and $.8<ICC\le 1$ (near perfect) as suggested by \cite{Telesford2010}.


\section{}\label{tab:appendix}
\renewcommand{\thetable}{\Alph{section}.\arabic{table}}

\begin{table}[ht!]
\centering
\caption{\small Performance of siGGM and SC naive approaches on simulated network data with p=100 and 200 nodes. Eglob is the bias in global efficiency.}
\label{table:vsScmethods}
\begin{adjustbox}{width=1\textwidth}
\begin{tabular}{llllllllll}
\toprule\hline \\[-.7em]
 & \multicolumn{4}{c}{p=100} &  & \multicolumn{4}{c}{p=200} \\
\hline
 & Eglob & MCC & AUC & L1 &  & Eglob & MCC & AUC & L1 \\
\textbf{Small World} &  &  &  &  &  &  &  &  &  \\
Glasso  & 0.177 & 0.327 & 0.827 & 0.575 & & 0.128 & 0.333 & 0.757 & 0.668 \\
Space   & -0.206 & 0.585 & 0.839 & 0.407 & & -0.374 & 0.597 & 0.763 & 0.430 \\
siGGM($\eta=$0)  & 0.061 & 0.538 & 0.847 & 0.509 & & -0.019 & 0.506 & 0.843 & 0.587 \\
siGGM & 0.078 & 0.590 & 0.884 & 0.478 &  & 0.121 & 0.526 & 0.906 & 0.532 \\
\textbf{Scale Free} &  &  &   &  &  &  &  &  &   \\
Glasso  & 0.117 & 0.365 & 0.798 & 0.560 & & 0.038 & 0.324 & 0.657 & 0.605 \\
Space  & -0.219 & 0.495 & 0.772 & 0.491 & & -0.403 & 0.358 & 0.664 & 0.555 \\
siGGM($\eta=$0) & 0.005 & 0.509 & 0.808 & 0.528 & & -0.100 & 0.411 & 0.769 & 0.573 \\
siGGM      & 0.054 & 0.562 & 0.853 & 0.428 &  & -0.075 & 0.469 & 0.868 & 0.442 \\
\textbf{Erdos Renyi} &  &   &  &  &  &  &  &  &   \\
Glasso  & 0.245 & 0.247 & 0.789 & 0.859 &  & 0.065 & 0.182 & 0.659 & 0.837 \\
Space   & -0.125 & 0.580 & 0.824 & 0.465 &  & -0.415 & 0.253 & 0.638 & 0.577 \\
siGGM($\eta=$0) & 0.020 & 0.363 & 0.792 & 0.679 & & -0.204 & 0.208 & 0.661 & 0.700 \\
siGGM    & 0.124 & 0.442 & 0.861 & 0.624 &  & 0.049 & 0.514 & 0.862 & 0.689
\\ \\[-.3cm] \hline
               \bottomrule
\end{tabular}
\end{adjustbox}
\end{table}

\begin{table}[h!]
\centering
\caption{\small Performance of SC informed methods on simulated network data with p=100 and 200 nodes.  Eglob is the bias in global efficiency. }
\label{table:misspecify}
\begin{adjustbox}{width=.9\textwidth, height=.6\columnwidth}
\begin{tabular}{lccccccccc}
\hline
 & \multicolumn{4}{c}{p=100} &  & \multicolumn{4}{c}{p=200} \\
 & Eglob& MCC & AUC & L1 & & Eglob & MCC & AUC & L1 \\
\textbf{Small World} &  &  &  &  &  &  &  &  &  \\
G-Wishart, MI(a) & 0.120 & 0.592 & 0.698 & 1.345 &  & 0.175 & 0.468 & 0.865 & 1.510 \\
G-Wishart, MI(b) & 0.173 & 0.447 & 0.574 & 1.498 &  & 0.224 & 0.337 & 0.819 & 1.801 \\
G-Wishart, MII(a) & 0.117 & 0.567 & 0.676 & 1.319 & & 0.174 & 0.447 & 0.842 & 1.506 \\
G-Wishart, MII(b) & 0.171 & 0.424 & 0.650 & 1.492 & & 0.224 & 0.318 & 0.797 & 1.788 \\
\hline
aGlasso, MI(a)  & -0.145 & 0.522 & 0.903 & 0.724 & & -0.234 & 0.454 & 0.800 & 0.773 \\
aGlasso, MI(b)  & -0.240 & 0.407 & 0.869 & 0.782 & & -0.258 & 0.396 & 0.788 & 0.801 \\
aGlasso, MII(a) & 0.183 & 0.477 & 0.889 & 0.720 &  & -0.221 & 0.439 & 0.735 & 0.770 \\
aGlasso, MII(b) & -0.279 & 0.364 & 0.840 & 0.805 & & -0.272 & 0.376 & 0.775 & 0.814 \\
\hline
siGGM, MI(a)  & 0.078 & 0.590 & 0.889 & 0.478 &  & 0.121 & 0.526 & 0.906 & 0.532 \\
siGGM, MI(b)  & 0.112 & 0.500 & 0.880 & 0.531 &  & 0.166 & 0.419 & 0.875 & 0.603 \\
siGGM, MII(a) & 0.075 & 0.576 & 0.879 & 0.486 &  & 0.125 & 0.514 & 0.901 & 0.563 \\
siGGM, MII(b) & 0.122 & 0.490 & 0.846 & 0.547 &  & 0.169 & 0.406 & 0.870 & 0.633 \\
\hline
\textbf{Scale Free} &  &  &  &  &  &  &  &  &  \\
G-Wishart MI(a) & 0.104 & 0.590 & 0.695 & 1.410 &  & 0.146 & 0.471 & 0.864 & 1.503 \\
G-Wishart MI(b)  & 0.159 & 0.446 & 0.671 & 1.483 &  & 0.196 & 0.339 & 0.822 & 1.583 \\
G-Wishart MII(a) & 0.102 & 0.568 & 0.675 & 1.397 &  & 0.145 & 0.452 & 0.846 & 1.442 \\
G-Wishart MII(b) & 0.156 & 0.424 & 0.650 & 1.481 &  & 0.196 & 0.323 & 0.801 & 1.604 \\
\hline
aGlasso MI(a)  & 0.223 & 0.509 & 0.905 & 0.719 &  & -0.311 & 0.375 & 0.701 & 0.730 \\
aGlasso MI(b) & -0.195 & 0.460 & 0.864 & 0.697 &  & -0.312 & 0.336 & 0.685 & 0.736 \\
aGlasso MII(a) & -0.256 & 0.442 & 0.901 & 0.739 &  & -0.333 & 0.351 & 0.690 & 0.739 \\
aGlasso MII(b) & -0.252 & 0.404 & 0.877 & 0.740 &  & -0.303 & 0.320 & 0.658 & 0.746 \\
\hline
siGGM MI(a)  & 0.054 & 0.562 & 0.853 & 0.428 &  & -0.075 & 0.473 & 0.868 & 0.442 \\
siGGM MI(b)  & 0.093 & 0.467 & 0.822 & 0.457 &  & 0.131 & 0.359 & 0.843 & 0.492 \\
siGGM MII(a)  & 0.061 & 0.552 & 0.845 & 0.447 &  & 0.078 & 0.459 & 0.865 & 0.457 \\
siGGM MII(b) & 0.099 & 0.451 & 0.812 & 0.469 &   & 0.132 & 0.346 & 0.839 & 0.523 \\
\hline
\textbf{Erdos Renyi} &  &  &  &  &  &  &  &  &  \\
G-Wishart MI(a)  & 0.174 & 0.505 & 0.821 & 1.300 &  & 0.140 & 0.519 & 0.860 & 1.572 \\
G-Wishart MI(b)  & 0.239 & 0.368 & 0.765 & 1.491 &  & 0.187 & 0.380 & 0.821 & 1.976 \\
G-Wishart MII(a) & 0.171 & 0.483 & 0.807 & 1.277 &  & 0.139 & 0.501 & 0.838 & 1.560 \\
G-Wishart MII(b) & 0.237 & 0.349 & 0.747 & 1.464 &  & 0.186 & 0.365 & 0.805 & 1.956 \\
\hline
aGlasso MI(a) & -0.335 & 0.404 & 0.810 & 0.741 &  & -0.424 & 0.171 & 0.596 & 0.712 \\
aGlasso MI(b) & -0.339 & 0.353 & 0.803 & 0.761 &  & -0.423 & 0.176 & 0.600 & 0.708 \\
aGlasso MII(a) & -0.362 & 0.333 & 0.824 & 0.764 &  & -0.424 & 0.162 & 0.631 & 0.707 \\
aGlasso MII(b) & -0.360 & 0.318 & 0.815 & 0.759 &  & -0.424 & 0.147 & 0.648 & 0.709 \\
\hline
siGGM MI(a) & 0.124 & 0.442 & 0.861 & 0.624 &  & 0.049 & 0.514 & 0.862 & 0.689 \\
siGGM MI(b) & 0.186 & 0.363 & 0.838 & 0.646 &  & 0.110 & 0.380 & 0.826 & 0.697 \\
siGGM MII(a) & 0.122 & 0.421 & 0.852 & 0.638 & & 0.050 & 0.499 & 0.846 & 0.690 \\
siGGM MII(b) & 0.171 & 0.344 & 0.825 & 0.665 & & 0.106 & 0.367 & 0.810 & 0.706
\\ \\[-.3cm] \hline
               \bottomrule
\end{tabular}
       \end{adjustbox}
\end{table}

\begin{table}[h!]
\centering
\caption{Within- and between- module differences in functional connectivity between males and females. Bolded values with an asterisk indicate statistically significant modules at the .05 level of significance (FDR correction for multiplicity).}
\label{table:permwithin}
\begin{tabular}{lcccccccc}
\hline \hline \\[-.3cm]
\multicolumn{9}{c}{\textbf{\large Pre-Teen}} \\ \hline
 & Unknown & Visual & DMN & SM & Aud & EC & FP left & FP right \\ \hline
Unknown & 0 &  &  &  &  &  &  &  \\
Visual & 8 & 24 &  &  &  &  &  &  \\
DMN & 7 & 15 & 0 &  &  &  &  &  \\
SM & 3 & 14 & 3 & 2 &  &  &  &  \\
Aud & 2 & 17 & 13 & 5 & 4 &  &  &  \\
EC & 6 & 32 & 9 & 20 & 21 & \textbf{42$^*$} &  &  \\
FP left & 2 & 14 & 8 & 4 & 10 & 13 & 0 &  \\
FP right & 5 & 18 & 9 & 11 & 11 & 16 & 3 & 6 \\ \hline \hline \\[-.3cm]
\multicolumn{9}{c}{\textbf{\large Teen}} \\ \hline
 & Unknown & Visual & DMN & SM & Aud & EC & FP left & FP right \\ \hline
Unknown & 2 &  &  &  &  &  &  &  \\
Visual & 6 & 22 &  &  &  &  &  &  \\
DMN & 3 & 8 & 2 &  &  &  &  &  \\
SM & 1 & 11 & 9 & 4 &  &  &  &  \\
Aud & 3 & 7 & 7 & 13 & 4 &  &  &  \\
EC & 8 & 22 & 13 & 12 & 12 & 18 &  &  \\
FP left & 3 & 16 & 7 & 6 & 9 & 13 & 0 &  \\
FP right & 2 & 12 & 9 & 5 & 15 & 18 & 8 & 2 \\ \hline \hline \\[-.3cm]
\multicolumn{9}{c}{\textbf{\large Young Adult}} \\ \hline
 & Unknown & Visual & DMN & SM & Aud & EC & FP left & FP right \\ \hline
Unknown & 2 &  &  &  &  &  &  &  \\
Visual & 1 & 10 &  &  &  &  &  &  \\
DMN & 1 & 14 & 2 &  &  &  &  &  \\
SM & 1 & 10 & 9 & 4 &  &  &  &  \\
Aud & 1 & 15 & 5 & 5 & 8 &  &  &  \\
EC & 5 & 16 & 10 & 12 & 12 & \textbf{26$^*$} &  &  \\
FP left & 1 & 7 & 5 & 7 & 11 & 9 & 6 &  \\
FP right & 1 & 8 & 7 & 9 & 7 & 17 & 8 & 0\\ \hline \hline
\end{tabular}
\end{table}
\end{appendices}

\clearpage
\begin{spacing}{0.9}
\bibliography{myref}
\end{spacing}

  M.A.(2014), `Structurally-informed bayesian functional connectivity analysis', \underline{NeuroImage} {\bf 86}, pp. 294-305.

\end{document}